\documentclass{acm_proc_article-sp}
\usepackage[english]{babel}
\usepackage[autostyle]{csquotes}
\usepackage{graphicx}
\usepackage{color}
\usepackage{fancyhdr}
\begin{document}
\newcommand{\comment}[1]{  }
\conferenceinfo{}{Bloomberg Data for Good Exchange 2016, NY, USA}

\title{Managing travel demand: Location recommendation for system efficiency based on mobile phone data}

\numberofauthors{5}
\author{
\alignauthor
Yan Leng\\
       \affaddr{MIT Media Lab}\\
       \affaddr{Cambridge, MA}\\
       \email{yleng@mit.edu}
\and
\alignauthor
Larry Rudolph\\
       \affaddr{Two Sigma LP \& MIT}\\
       \affaddr{New York, NY }\\
       \email{larry.rudolph@twosigma.com}
\and
\alignauthor Alex 'Sandy' Pentland\\
       \affaddr{MIT Media Lab}\\
       \affaddr{Cambridge, MA}\\
       \email{pentland@mit.edu}
\and
\alignauthor Jinhua Zhao\\
	\affaddr{MIT}\\
	\affaddr{Cambridge, MA}\\
	\email{jinhua@mit.edu}
\and
	\alignauthor 
	Haris N. Koutsopolous\\
	\affaddr{Northeastern University}\\
	\affaddr{Boston, MA}\\
	\email{haris@mit.edu}
}\maketitle

\begin{abstract}
	
Growth in leisure travel has become increasingly significat economically, socially, and environmentally. However, flexible but uncoordinated travel behaviors exacerbate traffic congestion. Mobile phone records not only reveal human mobility patterns, but also enable us to manage travel demand for system efficiency. In this paper, we propose a location recommendation system that infers personal preferences while accounting for constraints imposed by road capacity in order to manage travel demand. We first infer unobserved preferences using a machine learning technique from phone records. We then formulate an optimization method to improve system efficiency. Coupling mobile phone data with traffic counts and road network infrastructures collected in Andorra, this study shows that uncoordinated travel behaviors lead to longer average travel delay, implying the opportunities in managing travel demand by collective decisions. The interplay between congestion relief and overall satisfied location preferences observed in extensive simulations indicate that moderate sacrifices of individual utility lead to significant travel time savings. Specifically, the results show that under full compliance rate, travel delay fell by 52\% at a cost of 31\% less satisfcation. Under 60\% compliance rate, 41\% travel delay is saved with a 17\% reduction in satisfaction. This paper highlights the effectiveness of the synergy among collective behaviors in increasing system efficiency. 
 
\end{abstract}

%Finally, we analyze the interplay between congestion relief and overall satisfied location preferences to show that manipulating travelers choice bundles can be used as a effective tool to achieve significant travel time savings. 

%\\category{H.4}{Information Systems Applications}{Miscellaneous}
%A category including the fourth, optional field follows...
%\category{D.2.8}{Software Engineering}{Metrics}[complexity measures, performance measures]

%\category{H.4}{Information Systems Applications}{Miscellaneous TODO}

%\terms{TODO}


\section{Introduction}
Careful coordination of the travel behavior of tourists may reduce congestion, improve their travel experience, improve the quality of the environment, improve the quality of life for the local population, and avoid the many problems with tourists \cite{ivanovic2009fresh, de2015personalized, ccolak2016understanding}. 
Coordination has grown to become even more important as leisure travel has continued to increase, both internationally and domestically, contributing to 10\% of global GDP and 6\% exports in 2015 \cite{unwto}. International tourism increases around 4.2\% to 6.6\% since 2010 and reached a record of 1.2 billion traveler-arrivels in the same year \cite{unwto}.  This paper proposes a personalized event recommendation system that mitigates the adverse effects of mass tourism, while respecting the needs of all stakeholders. 

\begin{figure}
		\label{flex}
		\begin{center}
		\includegraphics[width=.8\linewidth]{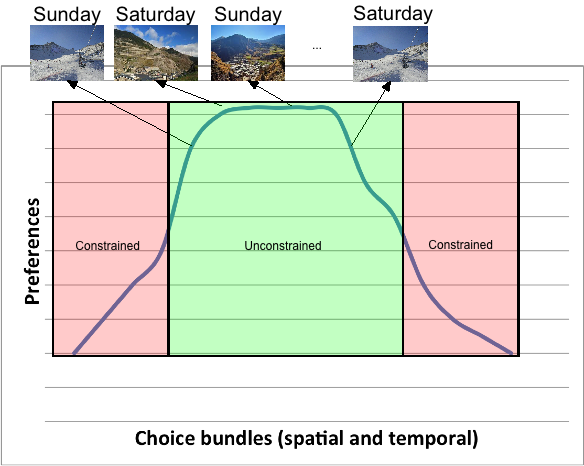}
		\end{center}
	\caption{Choice flexibilities.}
	\end{figure}

Our solution can be considered a type of {\em travel demand management}, i.e., a paradigm to reduce or shift travel demand in space or time to reduce the negative impacts\cite{halvorsen2015improving}. However, little research has focused on managing tourism travel demand. Unlike commuting trips, where the travel time and desinations are fixed, leisure travelers are more flexible. Figure \ref{flex} illustrates how individuals' preferences of location and time bundles are relatively flat above a certain threshold, indicating the flexibility of tourists' travel \cite{de2015personalized}. 
%This flexibility enables individuals to collectively make better decisions at a systematic level by sacrificing little individual benefits.

%Another critical component enabling location recommendations for system efficiency is the possibility in exploiting personal choice flexibilities at activity, location and temporal dimensions, specifically, what to do, where to go and when to go. Some travelers, especially for tourism purposes, care more about the activity than the locations for conducting the activities. This is more common for leisure purposes. For these groups of tourists, giving them recommendations on where and when to go would help them make informed decisions. Moreover, on the temporal side, when travelers make decisions on when to carry out an activity, they will make decisions blindly within some constraint if no information is given \cite{de2015personalized}. Many travelers want to minimize the amount of congestion experienced on the routes and are willing to change destinations if heavy traffic is expected. The time for traveling is flexible under certain time constraint. Under these conditions, which is often the cases, the visibility of what other people's behaviors and congestions along road links will guide them make better decisions. 

Recommendation systems have been successful tools in eCommerce. Current location recommendation algorithms, mostly adopted from movie or book recommendations, simply make recommendations based on inferred  personal preferences. Applied to travel, however, this method may lead to even more severe congestion and longer waiting times if the most popular locations are recommended at the same time. We argue that \emph{system efficiency}, the interplay between the location preferences at a user level and the traffic congestion at a system level, should be balanced when using location recommendations, which can serve as strategies in travel demand management. 

\comment{need to point out the it is the whole experience that is important which includes getting to the event.  That the tradeoff does not mean great sacrifice on the part of the tourist.}

The availability of large-scale geolocation data from mobile devices, such as the Call Detail Records (CDR) used in this study, offers an unprecedented opportunity for location-based service providers, transportation agencies, tourism departments, and governments to understand human mobility pattern, provide personalized information and improve system-wide performance \cite{brockmann2006scaling, gonzalez2008understanding, song2010modelling}. Making the comprehensive picture of population-wide behaviors enables decision-makers to intervene at a system level.
Call Detail Records, used to understand travel behaviors, have rarely been used to understand and manage travel demand or recommend locations \cite{de2015personalized, jiangactivity=, Alex2015}. 
Therefore, a recommendation system can be built based upon not only satisfying personal preferences at the demand side but also making the best use of the system capacity at the supply side.

In this paper, we use location recommendation to manage travel demand to achieve system efficiency. We show that uncoordinated travel behaviors lead to severe traffic congestion. At the same time, we propose a method as a solution for transportation practitioners or authorities to optimize and trade-off satisfied preferences and road congestion. We use matrix factorization to mine travelers' implicit preferences, taking advantages of underlying similarities among locations and travelers. We then formulate an optimization problem to maximize satisfied location preferences at the user level under pre-defined road congestion constraints. The method reveals the interplay between system congestions and user preferences. With an implementation with the CDR data in Andorra under various compliance rates, we show the effectiveness of the method. For example, under a 100\% compliance rate, a 52\% reduction in travel delay (from 11.73 minutes per hour to 5.61 minutes per hour) only sacrifices 31\% satisfaction regarding the recommendations. Similarly, under a 60\% compliance rate, a 41\% reduction in travel delay (to 6.98 minutes per hour) only sacrifices 17\% in satisfaction regarding the recommendations.

This paper is organized as follows: Section \ref{work} summarizes current literature regarding travel demand management and location recommendations. The data required to perform the study is described in section \ref{data}. Section \ref{method} demonstrates the framework of the method and details each step, including preference inference, and collective satisfaction maximization. Section \ref{case} describes a case study in Andorra. The performance of the proposed method is compared with a baseline model where location is recommended based simply on travelers' preferences. The impact of the method is analyzed under various compliance rates. Section \ref{future} concludes the paper and discuss future work. 

\section{Related works}
\label{work}
\subsection{Travel demand management}

Travel Demand Management (TDM) encompasses strategies that alter demand patterns to increase transportation system efficiency, instead of adding more capacity to the system \cite{halvorsen2015improving}. TDM has broad applications, including energy savings, air quality improvements, peak period congestion alleviation, etc \cite{tdm}. 

Different categories of TDM strategies include, economic policies, physical change measures, legal policies, and information or education measures \cite{garling2007travel, halvorsen2015improving}. Economic policies, the most popular TDM strategies, include taxing vehicles, congestion pricing, lowering transit costs, etc \cite{santos2005urban, halvorsen2015improving}. Physical change measures, such as walking/cycling improvements, park-and-ride schemas, represents another category in TDM \cite{tdm_pr}. Legal policies, prohibit traffic in some areas or parking controls. 

Many of the above strategies are not applicable in the context of tourism. Relatively little TDM research has targeted tourism demand, which is more flexible than commuting-related travels. Therefore, the following characteristics of this work differentiate it from prior travel demand management researches, namely: 
\begin{itemize}
	\item We focus on flexible travel demand, which can be manipulated at the destination and time-of-day levels.
	\item We propose to use Call Detail Records, a large-scale and opportunistic data source, to understand travel patterns.   
\end{itemize}

\subsection{Location recommendations}
In the early 2010s, several studies introduced traditional recommender engines to personalized location recommendation. Ye (2011) \cite{ye2011exploiting} introduced user-based and item-based Collaborative Filtering (CF) to location recommendations using user check-in data, based on the assumption that similar users have similar tastes and users are interested in similar Points of interests (POIs). Berjani (2011) \cite{berjani2011recommendation} employed the more effective and efficient matrix factorization in POI recommendations on check-in history. Regularized matrix factorization is used to learn the latent feature matrix, which has better perfornance than item-based CF. Recent research focuses on utilizing additional information. Geospatial factors, social networks, and temporal influences are three main examples. Other researchers argue that users prefer nearby locations rather than distant ones, which is defined as the geographical clustering phenomenon \cite{Zhang:2013:IPG:2525314.2525339,  yuan2013time}. Some researchers make the assumption that friends share more common interests than non-friends to utilize social influence in recommendation \cite{Pham:jucs_17_4:a_clustering_approach_for, Gao:2014:PLR:2645710.2645776}. Finally, to make use of temporal influence on activities, some researchers make seperate location recommendations for different temporal states \cite{Bao2015, yuan2013time, lian2014geomf}. Quercia (2010) \cite{l5694070} is the only work that makes recommendations using mobile phone data. However, his paper is based exclusively on item-based CF, which is computationally inefficient and hard to scale \cite{ye2011exploiting, wang2013location}. Application of existing methods, ignoring service capacity constraints, may result in traffic congestion and long waiting times, no matter how sophisticated these methods are in inferring preferences.

Therefore, this paper has the following improvements in location recommendation: 
\begin{itemize}
	\item We argue that capacity constraints are an important characteristic of location recommendation, which is currently ignored by the literature. We integrate capacity constraints in the location recommendation method. 
	\item We develop a framework to recommend locations for system efficiency based on Call Detail Records. It can be used in other cities when call records, traffic counts and road network GIS files are available. It can also be applied on other longitudinal data sources, such as WiFi, GPS, AFC, etc.   
\end{itemize}

\subsection{Next-location prediction}

Next-location prediction has been an increasingly popular topic in pervasive computing based on GPS, bluetooth, check-in histories, etc. Accurate next-location predictions, given previous footprints from these data sources, is a significant building block benefiting many areas, including mobile advertising, public transit planning, and urban infrastructure management \cite{alhasouncity, gomes2013will, lu2013approaching, petzold2006comparison}. Different data sources vary at spatial and temporal scales, and depending on the availability of contextual information \cite{de2013interdependence}. Most researchers build Markov models and predict longitude and latitude as continuous variables based on previous travel trajectories \cite{asahara2011pedestrian, ashbrook2003using, gambs2012next}. Mathew \cite{mathew2012predicting} predicts next-location using a Hidden Markov Model with contextual information, such as activities and purposes. Domenico \cite{de2013interdependence} and Alhasoun \cite{alhasouncity} uses mobility correlations, either measured by social interactions or mutual information, to improve forecasting accuracy. Though extensive researches have acceptable accuracy in predicting next locations, the performance is poor with when Call Detail Records are sparse \cite{alhasouncity}. 

In this paper, we develop a new perspective in viewing mobility behaviors based on Call Detail Records as sentences. We then introduce the use of Recurrent Neural Network, a successful tool in language modeling, into mobility prediction.

%%%%%%%%%%%%
\section{Data}
%%%%%%%%%%%%
\label{data}
In addition to the Call Detail Records, we make use of the topology of the Andorra road network, its capacity, and periodically recorded traffic counts.   This information is combined to understand travel demand patterns and transportation system performance.

\subsection{Call Detail Records}

\begin{figure}
	\centering
	\includegraphics[width=1\linewidth]{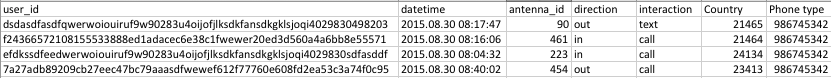} 	
	\caption{Snapshot of Call Detail Records.}
	\label{cdr}
\end{figure}

The Call Detail Records were originally used for billing purposes. A record is stored when a mobile phone user connects to the network of mobile carriers. Each Call Detail Record entry contains an encrypted user ID, start and end times of the phone call, the IDs of the connected cell tower ID, and the origin nationality and model type of the phone,  see Figure \ref{cdr}. The cell tower ID is easily converted to geographic location of the cell tower in the provider's network.  

\begin{figure}
		\centering
		\includegraphics[width=1\linewidth]{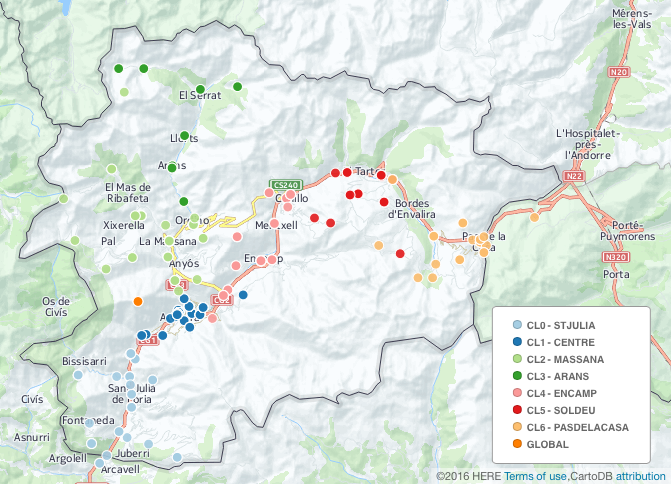} 
		\caption{Cell tower distributions in Andorra. \normalfont{Each circle represents a cell tower. Cell towers in the same city are colored the same, with the legend an abbreviation of a city name.}}
		\label{tower}
	\end{figure}

The anonymized CDR data used in this paper is collected from Andorra, a small country situated between France and Spain. As a case study, we target travelers visiting Andorra during a week in May, 2015 to provide personalized recommendations for system efficiency. These 47743 tourists include 20311 tourists from France and 27432 from Spain. The spatial distribution of the cell towers are shown in Figure \ref{tower}, with different colors representing different cities in which the tower is located. 

\subsection{Traffic flow derived from CDR and traffic counts}

\comment{Please check this for correctness}

We assume that the number of travellers using their cell phones while they on the highway is a constant fraction of the number of vehicles on the highway. Under this assumption, traffic flow  derived from the Call Detail Records can be used to approximate actual traffic flow.  

The Andorra road network is limited with a single major route between each city. Traffic counts have been collected at key locations using cameras by local authorities to monitor internal mobility. This data is publicly available \cite{andorra_traffic}.
%The monthly traffic counts of six key locations are selected as the ground truth to scale up CDR-based traffic flows to actual traffic flows. 
The GIS shapefiles of road networks were acquired from the Andorra Transportation Department. Important attributes used in the simulation include connecting cities, number of lanes, capacities and free flow travel time, which enable the computation of travel times based on traffic flow. Free-flow travel time were obtained from the Google Map API. Road capacities were obtained from a NCHRP report \cite{report_road}. 

From the phone record time and cell tower information, we can generate an Origin-Destination matrix, the number of trips between cell towers, i.e. Combining it with the road network information gives traffic flow over time, by group, and between two locations.

%	\begin{figure}
%		\centering
%		\includegraphics[width = 1\linewidth]{Pics/road.png} 	
%		\caption{Andorra road network skeleton. \normalfont{The red circle represents each city, connected by road links colored in blue. The travel time under free flow condition is labeled on the link.}}
%		\label{road}		
%	\end{figure}
	
\section{Method}

\label{method}

\begin{figure}
	\begin{center}
		\includegraphics[width=1\linewidth]{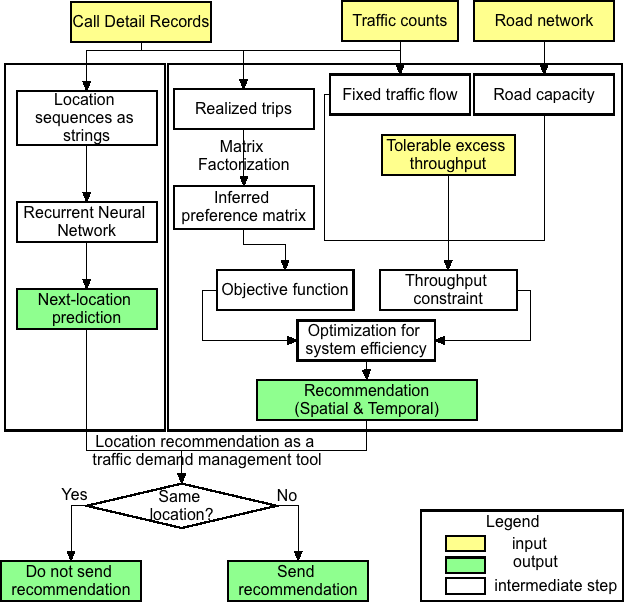}
		\label{frame}
	\end{center}

	\caption{Methodological framework. }
%\normalfont{Rectangle in yellow, green and white each represents input, output and intermedia step respectively. The method consists of four main steps, inferring traffic flow, inferring preferences, optimization for system efficiency and predict next-location. }
\end{figure}

The goal is to send recommendations for places to visit and when to visit them.
This is done in three steps, which are outlined in Figure \ref{frame}. 
The first step infers travel demand in terms of vehicle trips along road links, from mobile phone records. 
The second step infers personal location preferences based on location traces with no explicit ratings. With matrix factorization, we infer these implicit location preferences regarding all the locations with hidden factors, which correlate with the characteristics of both travelers and locations. With the inferred preferences, an optimization problem is formulated with the objective of maximizing satisfactions regarding the recommendations subject to tolerable congestion levels, which can be determined by the decision-makers. 
Finally, we predict next locations based on historical traces and compare them with the recommended locations to determine whether a recommendation will be sent. 
\comment{I do not understand this last step.}

\subsection{Terminology}

The following terms are used throughout the method as defined below. \\
\textbf{User profile}. User profiles contain the longitudes, latitudes, timestamps, and characteristics of the user. A user profile $l_{u, g} $ is generated for each user based on individual mobility traces ($l_{u,t_{1}, g}$, $l_{u,t_{2}, g}$, ...). $g$ is the user group of user $u$ based on his/her characteristics directly obtained or inferred from CDR. $t_i$ is the number of presences at location $i$.\\
\textbf{Realized trips}. Realized trips $p_{ij}$ are calculated by the number of times individual $i$ traveled to location $j$. Realized trips are used as a proxy for location preferences. \\
\textbf{Idealized trips}. Idealized trips $\hat{p}_{ij}$ measure the preferece of traveler $i$ regarding the location $j$ with no observed presences. Similarly, it is a measure of proxy travelers' preferences regarding the locations with no observations. \\
%\item \emph{\textbf{Location preference}}. Location preferences (implicit) is measured by the realized trips to each location. 
\textbf{Tolerable excess throughput}. Tolerable excess throughput is determined by the tolerable congestions by the decision-makers.

\subsection{Traffic flow inference}
Mobile phone data provides an imperfect estimate of the travel demand and traffic flow along road links. In order to understand travel delays, CDR data need to be processed to derive individual movements and vehicle trips from
the actual flow which is derived from the
 CDR-based travel demand and the actual travel demand.

The first step is to extract the tower-to-tower Origin-Destination (O-D) matrix, aggregated by the individual movements from one cell tower to another cell tower. Peak hour traffic flow can be learned as traffic flow varies heterogenously by links and hours of the day. The O-D pairs are assigned to road links given the road network. The last step is to scale the aggregated movements to actual vehicle trips using traffic counts as the ground truth, which is calculated in Equation \ref{flow_2}. 

\begin{equation}
	\label{flow_1}
	R_{it} = \sum_{jkt}O_{jt}D_{kt}
\end{equation}

\begin{equation}
	\label{flow_2}
	TC_{it} = R_{it}\times \beta_{it}
\end{equation}

where $i$ is the index of road link, $t$ is the index for hour of day, $O_jD_k$ represents OD pair with origin $j$ and destination $k$, $R_i$ is the vehicle trips along road link $i$, $TC$ is the actual traffic counts and  $\beta$ is the scaling factor.  

\subsection{Preference inference }

As no explicit review or rating regarding locations is available in CDR, we propose to use ``visiting frequencies'' as a proxy for location preferences. The next problem is to infer preferences regarding locations with no observed visits. 

Matrix factorization, one type of latent factor model, is used to infer travelers\textquoteright{} preferences regarding new locations. This model characterizes both the locations and users by vector of factors inferred from location visiting patterns, mapping both travelers and locations to a joint latent factor space of dimensionality $k$. The latent factor space determines why and how traverlers like each location based on hidden characteristics, which can be interpreted as personal interests or land use categories. High correspondence between location and user factors, based on the characterization of both the locations and the travelers, leads to a recommendation \cite{lian2014geomf}. 

%Specifically, each user $u$ is characterized by a vector $p_{u}$ measuring the extent to which the tourist is interested in the locations. The dimension of $p_{u}$ is the number of hidden factors, the size is represented by $k$. The resulting dot product $l_{i}^{T}p_{u}$ captures the interaction between tourist $u$ and location $i$, as shown in Equation (\ref{mf0}). This predicts tourists' interests regarding locations with no observed visits. Equation (\ref{mf1}) is a representation of the corresponding matrix operations of (\ref{mf0}). 

%\begin{equation} \label{mf0}
%\hat{p}_{ij} = u_{i}^{T}l_{j} = %\sum_{k=1}^{K}u_{ik}^{T}l_{kj}
%\end{equation}

%\begin{equation} \label{mf1} 
%P \approx U \times L^{T} = \hat{P} 
%\end{equation} 

%where $P$ is the realized preference matrix and $\hat{P}$ is the idealized preference matrix. 

\begin{figure}
	\begin{center}
		\includegraphics[width=1\linewidth]{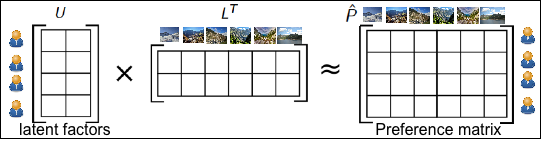}
	\end{center}
	\caption{Illustration of the preference inference methodology via matrix factorization to infer the preferences of travelers regarding locations with no observations. Matrix $U$ captures user's characteristics or interests  (hidden factors). Location matrix $L$ characterizes the associations of the locations with the latent factors (point of interest categories). The multiplication of the two predicts travelers' preferences across the location space.}
\end{figure}

\subsection{Optimization for system efficiency - individual preferences vs. congestion}

The key idea of the proposed method is to optimize travelers'  location preferences with the constraint of acceptable congestion. The authorities will then have the freedom to tradeoff between these two factors. An optimization model was built to maximize preferences regarding location recommendations subject to road capacity constraints. This model can be easily extended to other cases where capacity constraints exist by modifying the constraint accordingly. 

This paper formalizes the traffic problem by modeling destination and time choice as follows: each traveler $i$ makes a choice of location $j$ and the day for travel day $t$. The choice is made based on personal utility $p_{ijt}$, which is assumed to be the preferences regarding the location $j$ inferred from the call records. Since travelers make selfish choices independent of any other larger constants, the system may settle into a suboptimal state. In a suboptimal state, the travel time delay of the whole system as well as the congestion are higher than they should be. The set of destination choices that occur when every traveler maximizes their satisfied preferences is referred to as the user equilibrium flows \cite{de2015personalized, acemoglu2016informational}, which is similar to Wardrop's principles in route choice \cite{wardrop1952road}.

The objective function of the formulated optimization model aims to maximize the overall satisfied preferences regarding the location recommendations. The constraint function is determined by the acceptable congestion of the decision-makes.

\subsection{Next-location prediction with Recurrent Neural Network}

In order to distribute recommendations more efficiently, the method sends recommendations only when the predicted next-locations are not in line with the ones to be recommended. In this section, we predict individual next-locations based on historical location traces. 

Recurrent neural network is an adaptation of the traditional feed forward neural network, which can process variable-length sequences of inputs. It has been successfully used in language models, learning word embeddings, financial time series predictions \cite{pascanu2013construct, rumelhart1985learning}. In this project, we apply sophisticated recurrent neural network into mobility prediction by mapping between mobility models to language models. Cell tower traces for each individual are modeled as a sentence and each cell tower as a word. We use a simple RNN architecture, with a input layer, a hidden layer, a long short-term memory layer and output layer. The cell tower with the maximum probability is predicted to be the next location. The predicted next-location can be compared with the recommended location to determine the action to be taken. 

%RNN is applied to infer the real-valued representations of cell towers, relating "similar" cell towers close to each other in the vector space. 

RNN is advantageous in predicting next locations in two ways.\\
{\em Location sequences:} Travelers visit locations in a sequential way and RNN reads in data sequentially.\\
{\em Variable number of visited locations:} The heterogeneity in the size of location traces and frequency of mobile phone usage makes traditional machine learning techniques inapplicable. The ability to handle variable input lengths makes RNN appropriate in this situation \cite{de2015survey}.

\section{Application}
\label{case}
We performed extensive simulations to gain insights into the interplay between satisfaction regarding the recommendations and the travel delay caused by congestion. We vary the compliance rate across the population,  i.e. the probability that travelers will follow the recommendation, in order to evaluate the potential traffic improvement of the recommendation system.
The simulation assumes that the individuals who do not comply will follow their preference with no behavioral change. 

\subsection{Location recommendation for system efficiency}

Using the method described in section \ref{method}, we simulate the average travel delay per hour and overall idealized trips to assess the system-wide impacts and effectiveness of the method. The notion of idealized trips is a measure of satisfaction regarding the recommendtions. Average travel delay measures the congestion externalities of the recommendations. In transportation, practitioners and planners model the relationships between traffic flow and travel time based on the characteristics of the road infrastructure. One of the most widely used methods is Bureau of Public Road function (BPR) \cite{akcelik1991travel}, which models travel time as a function of the ratio between actual traffic volume and road capacity, volume-over-capacity (VOC) \cite{ben1999discrete}, as shown in Equation (\ref{bpr1}). 

\begin{equation}
\label{bpr1}
t_{\text{simulated}} = t_{\text{free-flow}} \times (1 + \alpha(V/C)^\beta)
\end{equation}

Average travel delay: 
\begin{equation}
\label{bpr2}
\Delta t = t_{\text{simulated}} - t_{\text{free flow}}
\end{equation}
where $t_{\text{free-flow}}$ is the free flow travel time on the road segment, $t_{\text{simulated}}$ is the simulated travel time, $\Delta{t}$ is the delay caused by congestion, $\alpha$ and $\beta$ are parameters that are used to characterize the non-linear relationship between $V/C$ and $t_{simulated}$. The default parameter for the BPR equation are $\alpha=0.15$ and $\beta=4$. 

\begin{figure}
	\begin{center}	
		\includegraphics[width=1\linewidth]{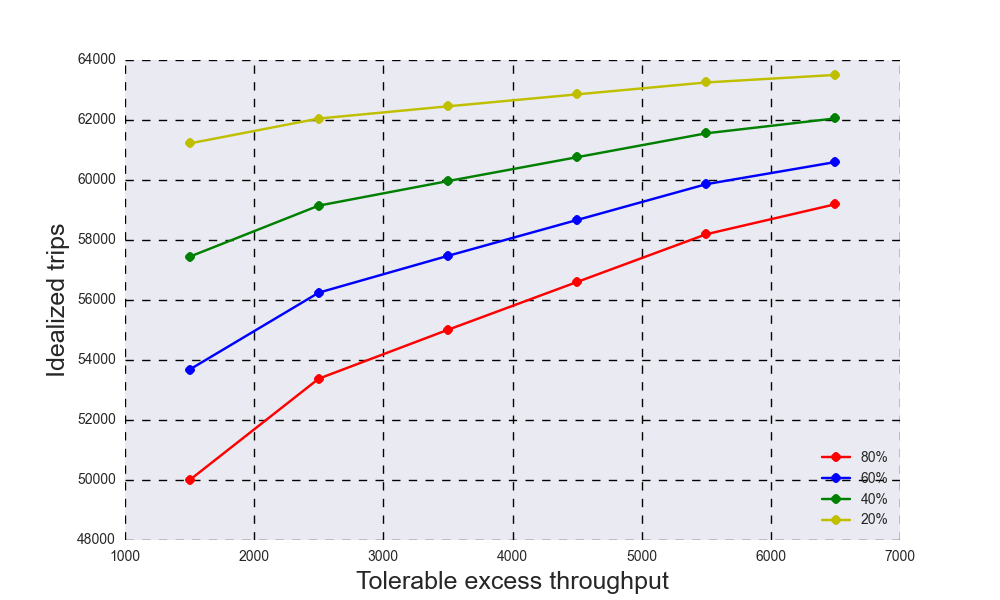}

	\end{center}
	\caption{Relationships between idealized trips and tolerable excess throughput.}
	\label{compl_ana1}
\end{figure}

\begin{figure}
	\begin{center}

		\includegraphics[width=1\linewidth]{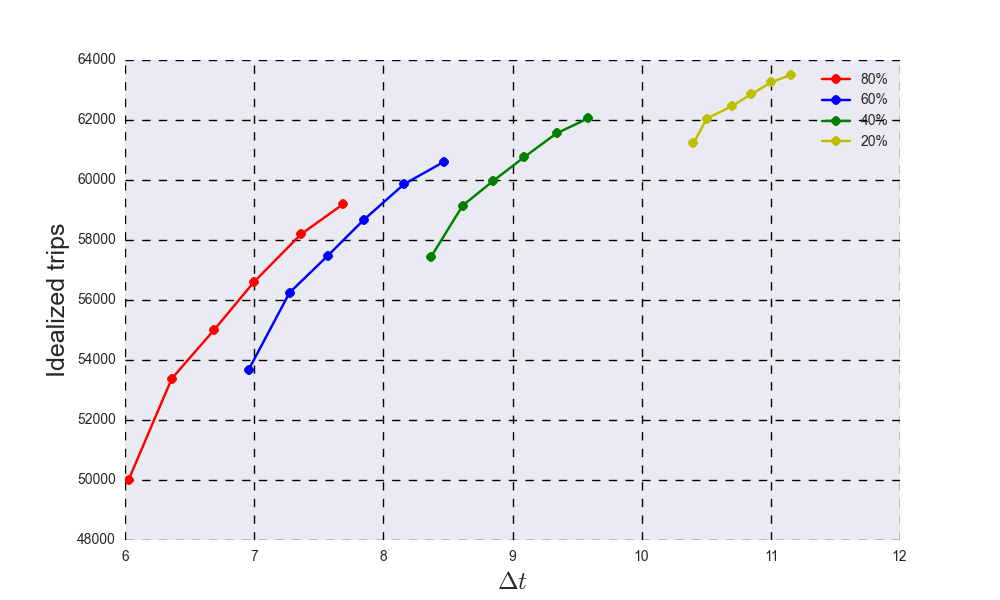}
	\end{center}
	\caption{Interplay between idealized trips and average travel delay.}
	\label{analysis_2}
\end{figure}

We compare our method with a baseline model, which is referred to as \emph{preference only} method. This method makes recommendations simply according to personal preferences with no system performances taken into account. Table (\ref{res_c}) summarizes average travel delay per hour and idealized trips of: 1) the preference only method; 2) the proposed method under various compliance rates. Under full compliance, the  average travel delay is 5.61 minutes per hour with 44930 idealized trips, which indicates that with a 52\% reduction in congestion time, only 31\% of idealized trips are sacrificed. When the compliance rate is 80\%, a 47\% reduction in congestion time is achieved with only 23\% of idealized trips being sacrificed. The lower the compliance rate, the larger the idealized trips and the longer travel delay. 
\begin{table}
		\begin{tabular}{p{2.5cm}|p{2cm}|p{2.5cm}}
			\hline
			\textbf{Scenario}  & \textbf{$\Delta \text{\textbf{t}}$ (min/h)} & \textbf{Idealized trips} \\ 			
			%Status quo   &     18.58                  &                 NA                            
			\hline
			Preference only  &              11.73                   &           64925              \\ \hline 
			100\% comp. &       5.61                  &     44930        \\ \hline

			80\% comp. &       6.17                  &     49997        \\ \hline			
			
			%			80\% compliance  &          6.17                  &     49997                  \\ 
			60\% comp.  &       6.98                  &     53680                  \\ \hline
			%			40\% compliance   &      8.37                  &     57442                  \\ 
			40\% comp.  &       8.37                  &     57442                  \\
			\hline			
			20\% comp.  &          10.40                  &     61219           \\
			\hline
		\end{tabular}
	\caption{Results comparisons. 
		\normalfont{$\Delta t$ is the average travel delay during the peak hour. Idealized trips measures travelers' satisfaction regarding the recommended locations. Comp. is short for compliance rate. Preference only is the baseline model where recommendations are based on predicted preferences.}}
		\label{res_c}
\end{table}
Figure \ref{compl_ana1} shows the relationship between idealized trips and tolerable excess throughput, which indicates individual perceived benefits under various level of congestion controls. Higher compliance rates satisfy larger individual benefits, which generate more traffic. The concave relationship reveals that preferences are satisfied more quickly in the beginning and slow down afterwards.  The tolerable excess throughput enables decision-makers to manage congestion at an acceptable level.
%\ref{analysis_2}
Figure \ref{analysis_2} reveals the interplay between idealized trips and average travel delay. A horizontal reference line shows that for the same level of idealized trips, synergized behaviors generate less congestion. On the other hand,  a vertical reference line shows that for the same travel delay, coordinated behaviors are more effective. This indicates the importance of effective schemas to incentivize behavioral change to achieve synergetic effects. 

%\normalfont{ Different compliance rates are represented in different colors as shown in the legend. This figure indicates that coordinated behavior is more efficient: 1) given the same travel delay, higher compliance rate is more satisfactory; 2) given the same level of satisfaction regarding the recommendations, higher compliance rates lead to shorter travel delay. }

\subsection{Next-location prediction}
As a case study, we evaluate different methods using CDR data collected over three weeks during May, 2015 in Andorra. We apply the method specifically on tourists, which can be filtered on the country code from CDR. We do not exclude travelers with too few observations as
long as their travel call includes more than one cell tower, suggesting the generalarization of the method. We use two settings with different spatial resolutions, at the cell-tower level and the merged -cell-tower level.

We introduce two baseline models: \enquote{Most Frequent} model and Markov model. \enquote{Most Frequent} model, referred to as the naive model, predicts the next location using the most frequent location. The Markov model is built on the contextual co-occurences between sequences of locations \cite{hsieh2015t, gambs2012next, lu2013approaching}. Various parameters have been tuned to optimize the performance of RNN, including activation function, dimension of embedding layer, drop out rates, sample size, and batch size. 

Tables \ref{compare} and \ref{compare_1} show that the naive model has 50\% and 63\% accuracies in predicting next location at the cell-tower and merged-cell-tower level. The Markov model across the population has 54\% accuracy at the cell tower level, with 8\% improvement. However, at the merged cell tower level, the population-wide markov model has lower accuracy than the naive model. Comparatively, RNN improves the accuracy of location prediction significantly, with an accuracy of 67\% and 78\% on the two settings. It improves 34\% and 41\% in accuracy compare with the two settings, indicating that RNN significantly improves the accuracy of the next-location prediction. 

	\begin{table}
		\caption{Cell-tower level} 
		\label{compare} 
		\begin{center}
			\begin{tabular}{|c|c|c|c|}
				 \hline
				& Accuracy & Improvement \\ \hline Naive model & 50\%  & NA \\ \hline Markov model  & 54\% & 8\% \\ \hline RNN & 67\% & 34\% \\ \hline
			\end{tabular} 
			\caption{Merged-cell-tower level} 
			\label{compare_1}			
			\begin{tabular}{|c|c|c|c} 
				\hline
				& Accuracy  & Improvement \\  \hline
				
				Naive model  & 63\% & NA \\ \hline Markov model  & 57\% &  -16\% \\  \hline RNN & 78\%  & 41\% \\ \hline
				
			\end{tabular}
			
		\end{center} 
	\end{table}

\section{Conclusions and Future work}
\label{future}
We have shown that individual travel decisions, without accounting for system efficiency, lead to traffic congestion. Existing location recommendation algorithms exacerbate congestion by recommending popular locations.  To address this issue, we develop a location recommendation method  to manage travel demand. Call Detail Records are the raw input and are processed to infer personal location preferences, traffic congestion, and other information about both tourists and native travellers. This study, as far as we know, is the first one making location recommendations based on Call Detail Records. Most importantly, we factor in the special characteristics, capacity constraints, of location recommendation, which distinguish our system from existing recommendation methods. The simultaneous trade-off between congestion relief and overall satisfied location preferences learned from the simulation results indicate that moderate sacrifices for individual utilities lead to significant collective travel time saving. 

The simulation results from our Andorra case study reveal a noticeable impact in reducing traffic congestion with moderate sacrifices on individual preferences. For example, under 100\% compliance, there is a 52\% reduction in travel delay (from 11.73 minutes per hour to 5.61 minutes per hour) with 31\% dissatisfaction rate regarding the recommendations. Even with a much smaller compliance rate,  under 60\%, there is a  41\% reduction in travel delay (to 6.98 minutes per hour)  with only a  17\% dissatisfaction rate. We use a recurrent Neural Network approach to make sense of the input data.  With the implementation of RNN on the large-scale CDR collected in Andorra, this paper demonstrates the applicability of the method with accuracies of 67\% and 78\% at cell-tower and merged-cell-tower levels, representing an improvement of greater than 30\% compared to two base-line models. 

The method developed in this paper specifically targets leisure travel, where travelers are relatively more flexible in the decision-making processes at spatial and temporal levels. We aim to divert travelers to their 
\enquote{secondary} choices at the location and the time to visit, while gaining more travel-time savings both for the individuals and for the society as a whole.

This research opens up multiple directions to advance in the topic of location recommendation for system efficiency using Call Detail Records. In this study, we only use travel delay as the systematic efficiency measure. However, other externality measures, such as air pollution or energy consumption, should also be factored in for comprehensive evaluations from the system side. In addition, except for time and destination, other interesting factors could be included in the choice bundle, such as travel routes, budgets, etc.  
	
The proof-of-concept experiments in this study demonstrate the effectiveness of our approach. The natural next step is to investigate how to incentivize users to sacrifice perceived benefits for better system performances. A comprehensive framework of the application in real situations, detailing the distribution channel, frequencies, target markets, needs to be studied further from a marketing perspective. 
	
An interesting extension of the paper is the information configurations for travelers, specifically, how to strategically present information accounting for travelers' willingness to accept the information. An web-interface-based experiment could be developed to help understand travel behaviors and decision-making processes, and how the system dynamics perform by providing various recommendation configurations. 
	
Call Detail Records constitute one longitudinal data source in understanding travel behaviors. Integrating with other data, such as WiFi and bluetooth data, could supplement it and enhance the application - providing better spatial and temporal resolutions, social media data - capturing momentary feelings.

\section{Acknowledgement}
This project is a collaboration with Changing Places Group at MIT Media lab, directed by Professor Kent Larson. We thank them for the data support and discussions.

\nocite{*}
\bibliographystyle{abbrv}
\bibliography{ref}

\end{document}